# High-power sub-picosecond filamentation at 1.03 μm with high repetition rates between 10 kHz and 100 kHz

Robin Löscher*[1], Victor Moreno[2], Dionysis Adamou[3], Denizhan K. Kesim[1], Malte C. Schroeder[1], Matteo Clerici[3,4], Jean-Pierre Wolf[2], Clara J. Saraceno[1]

[1])Photonics and Ultrafast Laser Science, Ruhr-Universität Bochum, 44801 Bochum, Germany

[2])Groupe de Physique Appliquée, Université de Genève, CH1211 Geneva, Switzerland

[3])James Watt School of Engineering, University of Glasgow, G12 8QQ, Glasgow, UK

[4])Dipartimento di Scienza e Alta Tecnologia, Università dell'Insubria, Via Valleggio 11, 22100, Como, Italy

(E-Mail: *robin.loescher@ruhr-uni-bochum.de)

**Filamentation has been extensively explored and is well understood at repetition rates < 1 kHz due to the typical availability of multi-mJ laser systems at moderate average power. The advent of high-power Yb-lasers opened new possibilities for filamentation research. However, so far, high average power Yb systems have mostly been explored to increase the driving pulse energy to several hundreds of mJ, and not at significantly higher repetition rates. In this paper, we study for the first time long filaments at unprecedented high repetition rates of 10 kHz, 40 kHz, and 100 kHz using a 500-W Yb-doped thin-disk amplifier driver operating with sub-700 fs pulses. We compare the filament length, density hole, and fluorescence at constant peak power but different repetition rates and find a strong dependence on filament length and density depletion with repetition rate. Our analysis reveals the emergence of a significant stationary density depletion at repetition rates of 40 kHz and 100 kHz. The corresponding reduction in breakdown threshold by increasing the laser repetition rate observed in our study signifies a promising avenue for enhancing the efficiency and reliability of electric discharge triggering in various scenarios. Using capacitive plasma probe measurements, we address the limitations of fluorescence imaging-based measurements and demonstrate a systematic underestimation of filament length. This work contributes to a deeper understanding of the interplay between laser repetition rates, filamentation, and heat-driven density depletion effects from high-repetition-rate high-power laser systems and will contribute to guide future research making use of filaments at high repetition rates.**

## I. Introduction

High-energy ultrashort laser pulses can reach peak powers that give access to the process of filamentation[1,2]. By exceeding the critical power for self-focusing[3], the beam size decreases while propagating through a transparent medium. For 1-μm sub-picosecond laser pulses, the corresponding threshold in ambient air is approximately 6 GW, based on the nonlinear refractive index measured by Schwarz *et al.*[4], which is commonly met by modern ultrafast laser sources. Self-focusing then results in corresponding intensities of tens of TW/cm², which accordingly lead to under-dense plasmas by optical field ionization, locally decreasing the refractive index of the medium[1,2]. Strong nonlinearities, at peak powers well above the critical power for self-focusing, thus lead to cascaded Kerr-focusing and plasma-defocusing, resulting in confined self-guided light structures that exceed the initial Rayleigh length of the pump beam[1,5] and can reach lengths of several hundreds of meters[6]. Due to intensity clamping, the core intensity of the filament remains on the order of $10^{13}$ W/cm² in air[7], still allowing for single ionization of the air molecules by a single high-energy picosecond pulse[8].

Scientific applications of filaments include temporal pulse compression[9–11], free-space optical telecommunication[12–14], guiding and triggering of electrical discharges[15–17] and lightning[18] and generation of higher harmonics[10] and broadband THz radiation[19–21], many of which would immensely benefit from higher repetition rates.

Traditionally, ultrafast Ti:Sapphire amplifiers at 800 nm were the workhorse used to generate filaments due to the requirements of high peak power mentioned above. Ti:Sa amplifiers can easily achieve the required peak powers; however, they are limited in average power to a few tens of watts; therefore, repetition rates < 1 kHz were typically employed. However, in the last decade, Yb-laser systems have immensely progressed, and much higher average power ultrafast laser systems up to several kilowatts have become available. Laser gain media in slab[22], fiber[23], and thin-disk[24] geometries enable applications in high-repetition-rate regimes (≥ 5 kHz) while still providing sufficient pulse energy for filamentation[25–27] are now available, which has, in turn, raised attention towards filamentation effects at shorter interpulse separation times[14,28–31]. However, so far, most efforts in the direction of using high average power Yb-lasers have focused on increasing the driving single pulse energy rather than exploring significantly higher repetition rate regimes, where pulse-to-pulse accumulation effects can become significant and affect filament formation. Filamentation with a high-energy high-average-power 1030-nm picosecond thin-disk laser has been studied with repetition rates of up to 2.5 kHz[32]. Exploring filamentation at higher repetition rates using such high average power lasers, opens the door for novel physics and applications in this area.

The energy deposited through the plasma generation during the filamentation process results in a localized heated channel with reduced gas density[1]. These heat-driven density depression channels were found to be the dominating effect for electric breakdown[33,34]. Following these findings, the comparatively slow time scale of the heat diffusion raised interest in spark gap experiments at higher laser repetition rates, where cumulative effects would come into play[30]. Recently, cumulative air-density depletion driven by sub-mJ pulses at tens of kHz repetition rates was shown to produce a quasi-stationary state of reduced gas density, affecting the electric discharge triggering potential[35]. However, in this experiment, the rather low pulse energies available at high repetition rates made it difficult to perform a thorough characterization of the filaments and draw conclusions about the underlying physical mechanisms.

In this study, we report on filamentation from sub-picosecond laser pulses from a commercial 500-W average power Yb-doped thin-disk regenerative amplifier operating at 10 kHz, 40 kHz, and 100 kHz repetition rate providing multi-mJ pulse energies at all repetition rates and up to 50 mJ at 10 kHz repetition rate. This is, to the best of our knowledge, a filamentation regime unexplored until now. We investigate the filamentation-induced localized air heating at 40 kHz and 100 kHz and find out that a quasi-stationary density depleted zone is established due to cumulative heating of the ambient air. Furthermore, we characterize the plasma channel length by capacitive plasma probe (CPP) measurements and estimate the relative spatial distribution of the plasma density along the propagation axis. Measurements of the electrical breakdown reveal a nonlinear response of the breakdown voltage in high-repetition-rate filaments.

## II. Experimental setup and methods

The experimental setup is shown in Fig. 1. The pump laser in our study is a commercial Ytterbium-doped thin-disk regenerative amplifier with 1.03 µm central wavelength at a maximum average power of 500 W. The system provides pulses at 10 kHz, 40 kHz, and 100 kHz with pulse durations of 680 fs, 660 fs, and 660 fs, and a collimated $TEM_{00}$ beam diameter of 14.9 mm, 15.2 mm, and 15.8 mm, respectively. The maximum corresponding pulse energies and peak powers are 50 mJ (69 GW), 12.5 mJ (18 GW), and 5 mJ (7 GW). At the maximum power and for 10 kHz, the system provides pulses of up to $P_{peak} \approx 11 P_{cr}$ for the ambient air in our laboratory. At 40 kHz and 100 kHz the maximum available peak power amounts to $3P_{cr}$ and $1P_{cr}$, respectively. An active beam stabilization was implemented before generating the filament to reduce beam pointing instabilities from the source. The pulse energy to generate the filaments at a given repetition rate was externally attenuated by a half-waveplate and a thin-film polarizer. The filament was generated in a focusing beam geometry by a concave 0-degree mirror ($f = 1000$ mm) for all repetition rates and all pulse energies. Considering Gaussian beam optics only, i.e., without filamentation, this configuration results in a Rayleigh length of 1 mm to 2 mm for all repetition rates.

### A. Dimensions of the filament

The filamentation length was characterized by two different methods: imaging of the plasma fluorescence[36] and CPP method[37,38]. Calibrated images of the visible fluorescence of ionized molecules provide a measure of the dimensions of the plasma channel. We defined the length of the filament by the relative $1/e^2$ intensity for each image. However, weak fluorescence from low-density plasma, established far away from the focal plane, disappears in the noise of the camera chip.

In order to circumvent this limitation and access the real filament length, we performed CPP measurements, that allow for spatially resolved detection of charged particles in the filamentation region by translating the electronic probe along the laser propagation axis. Two square copper electrodes ($a = 18$ mm) with 19 mm separation were symmetrically placed around the plasma channel and connected to a DC electric bias of -8 kV. In the plasma channel, a fast, typically ns decay, polarization-driven charge separation dominates the initial transient after the laser pulse arrival[38,39]. This fast signal was rejected by opting for a slow sampling rate (200 kSa/s) synchronous to the laser repetition rate. The voltage drop was measured over a 50 kΩ resistor in series. This configuration allowed us to measure the misalignment-insensitive positive ionic current only[40,41]. To compare our results with previous studies, the average collected ionic charge per pulse has been calculated from the currents. For our study, we define the filament length obtained by the ionic current measurement to be the relative $1/e^2$ distance from the position where the peak current is observed.

### B. Density depletion

To retrieve the density depletion by localized heating inside the filament, the refractive index change has been measured by collinearly counter-propagating a continuous-wave probe beam along the plasma channel axis and resolving the accumulated phase shift in a folded-wavefront interferometer[42]. To this end, a matching f-2f imaging telescope was implemented to image the focal plane of the pump-focusing concave mirror[43]. Time-averaged images of 5 and 13 pulses for 40 kHz and 100 kHz laser repetition rate, respectively, were recorded with a camera exposure time of 125 µs. The accumulated phase shift of the probe beam was obtained with a common two-dimensional phase-retrieval algorithm from the background-corrected probe beam images[44]. The retrieved phase-shift yields the density variation by

$$\frac{\Delta n}{n_0}(x,y) = \frac{\varphi(x,y)\lambda_{\mathrm{p}}}{2\pi L_{\mathrm{fil}} n_0}, \quad (1)$$

where $\lambda_{\mathrm{p}}$ is the wavelength of the probe beam, $L_{\mathrm{fil}}$ is the length of the filament, and $n_0$ is the ambient gas refractive index. We can use the Lorenz-Lorentz relation[45] to calculate the relative gas density change

$$\frac{\Delta\rho}{\rho_0} = \frac{(n_0(1+\Delta n))^2 - 1}{(n_0(1+\Delta n))^2 + 2} \cdot \frac{n_0^2 + 2}{n_0^2 - 1} - 1, \quad (2)$$

where $\Delta\rho/\rho_0$ is the relative gas density change and $\Delta n$ is the absolute refractive index change.

### C. Qualitative plasma fluorescence

In order to characterize the emission of the filaments and study eventual chemical accumulation effects, the fluorescence of the filaments has been measured by using a UV-VIS spectrometer with an infinity focus lens, imaging into the spectrometer fiber.

### D. Spark gap experiments

Two identical spherical electrodes with high-voltage DC bias separated by 30 mm were placed close to the filament to measure the electric breakdown voltage. The breakdown voltage was measured at all repetition rates in two cases, namely at constant pulse energy and at constant average power.

## III. Results

### A. Dimensions of the filament

Figure 2 compares the lengths of the plasma channels retrieved by CPP and fluorescence imaging in the case of 10 kHz repetition rate. For both methods, we extract the relative 1/e² total length. The measurements show that the fluorescence imaging is significantly less sensitive at lower peak powers and cannot resolve plasma channels below a threshold of 12 GW, thus significantly underestimating the plasma channel length. The presence of plasma is a visual indication of filamentation, and our measurements show fluorescence regions significantly exceeding the initial Rayleigh lengths. In the case shown here for 10 kHz and 50 mJ pulse energy, the filament length as measured by fluorescence (orange points) exceeds the Rayleigh range by 51 times, with a size of 65 mm (1/e²). A relevant byproduct of the optical field ionization, however, is the generation of free carriers. Notably, the sensitivity of this measurement allowed us to record an ionic current over a distance as long as 202 mm. We used this metric to compare the length of the filaments at different laser repetition rates. It is clear that, for low-density plasmas, the fluorescence signal disappears in the noise of the detector. Furthermore, we observe a saturation of the length due to multi-filamentation, occurring where $P_{\mathrm{peak}} \geq 10 P_{\mathrm{cr}}$ [32]. Our characterization of the lengths of the plasma channel using the CPP method, see Fig. 3b, shows a decrease at higher repetition rates when measured at constant average power, consistent with the lower peak power in each case. This shows that the contribution of peak power continues to dominate over the contribution of the stationary density reduction. However, at peak powers above critical power for all repetition rates, see data for 5 mJ in Fig. 3a, we observe an elongation of the filament by increasing the repetition rate. This elongation occurs due to the balancing of Kerr self-focusing, plasma defocusing, and additional defocusing by the density depletion[46], see section III.B. From the spatial distribution of the plasma density in Fig. 4, we see that a plasma channel is established without filamentation, hence below the critical power for self-focusing. At peak powers above the critical power, the balanced Kerr-focusing and plasma-defocusing elongate the distribution asymmetrically, indicating that we are indeed operating in the filamentation regime. Figure 5 reveals that for all repetition rates, the total collected ionic charge, as a relative measure of the plasma density[38], scales linearly with peak power, which indicates that the initial plasma density increases linearly with the available peak power and photon bath, respectively. The lower ionic charge measured at constant peak power for higher repetition rates visible in Fig. 5 confirms that cumulative effects in the ionic species are negligible, whereas cumulative effects in the thermal properties of the gas (reduced gas density by increasing repetition rates) are dominating, as presented in the following section.

### B. Density depletion

At 40 kHz and 100 kHz, the fringes in the interferometric images of the density hole were temporally and spatially stationary[35] and recorded by the CCD, integrating the probe beam imaged over 5 and 13 pump pulses, respectively. The gas density reached a quasi-stationary state which is evidenced by the fact that, despite operating in a pulsed regime, interference fringes of high visibility were observed while imaging over multiple pulses. However, at 10 kHz the CCD integrated the images over 1.25 laser pulses. At this low repetition rate, the deposited heat diffuses faster into the ambient medium than the pulse separation time and recovers the density hole. The subsequent pulse, acting on the recovered air, again induces a spike in the gas density[35]. In this case, no stationary density-depleted zone was established. Time-resolved techniques could reveal the heat diffusion on a ns time-scale[42] in a future experiment. Fig. 6 shows the spatially resolved averaged density depletion along the filament. Note that, due to the time-integrated nature of our measurement, no transient effects, such as acoustic waves or density increase in the vicinity of the filament, are visible. At 100 kHz, the maximum relative density depletion increased by a factor of 2 compared to that at 40 kHz. This correlates with our findings from the electrical discharge experiments in the following section. We would like to point out that a lower gas density change was predicted by previous simulations[35], where the authors used an initial energy deposition in the filament of 500 nJ, significantly lower than in our study. This reinforces the necessity of empirical studies on filamentation at high repetition rates and high peak powers, crucial for understanding the hydrodynamics of filamentation.

### C. Spark gap experiments

The results of the breakdown potential measurements are shown in Table 1. An increase in the repetition rate reduces the required electric field for spark gap triggering at equal pulse energies and peak powers, respectively. We observed a breakdown potential decrease of 60 % from 10 kHz to 100 kHz laser repetition rate. Note that in this configuration no natural breakdown in the absence of a filament was observed up to a potential of -25 kV, limited by the high-voltage supply. We observe a decrease of several tens of percent in the breakdown potential corresponding to the magnitude of the density depletion. In a high-repetition-rate regime, where the pulse separation times are shorter than the timescale for heat diffusion, we have shown that the cumulative nature of the heat deposition leads to a quasi-stationary depletion state. However, it has been shown that the gas density reaches a minimum on the time scale of a few hundred nanoseconds after the incident laser pulse[42]. When considering high-repetition rate regimes simulations have shown that in the quasi-stationary depletion state each laser pulse still causes rapidly decaying gas density depression[35]. These temporal minima of the gas density, cumulatively affected in their peak value, dominate the breakdown potential, following Paschen's law[17,34,47], and thus warrant further investigation of the rapid transient intra-pulse hydrodynamics. The comparison of the breakdown potential at equal average powers shows that the length of the filament, due to higher pulse energies at lower laser repetition rates, overcomes the cumulative density depletion as a dominant effect on the breakdown potential, see Fig. 5.

*Table 1: Electric breakdown potentials in kV between identical 30 mm separated spherical electrodes at constant pulse energy and constant average power for laser repetition rates of 10 kHz, 40 kHz, and 100 kHz in air.*

|  | 10 kHz | 40 kHz | 100 kHz |
| --- | --- | --- | --- |
| 5 mJ | -23.5 ± 0.3 | -16.0 ± 0.1 | -13.9 ± 0.2 |
| 250 W | -10.9 ± 0.2 | -14.6 ± 0.2 | -19.9 ± 0.2 |

### D. Qualitative plasma fluorescence

A strong emission of fluorescence lines by cumulated $O_2^-$ ions[30] in the range between 520 nm and 850 nm was not detected by the spectral measurements. This indicates that photodetachment at does not significantly contribute to the enhanced breakdown efficiency at the higher repetition rate and constant energy. Measurements in nitrogen-purged conditions showed an increase in the corresponding nitrogen fluorescence intensity, while the breakdown potential remained constant to ambient air conditions within our measurement sensitivity.

## IV. Conclusion

We perform a thorough characterization of filamentation of high-repetition-rate lasers at 10 kHz, 40 kHz, and 100 kHz. In contrast to a previous study[28], an elongation instead of shortening of the filament by increasing the repetition rate has been observed when the peak power is kept constant. This, however, is consistent with the elongation of filaments by preformed density holes[46]; whereas in our experiments, the stationary density depletion due to cumulative heating rather than a preformed density hole causes the elongation. A quasi-stationary density depletion has been measured at repetition rates of 40 kHz and 100 kHz. We have not observed self-guiding of subsequent pulses inside thermo-acoustic waveguides, which may occur at higher repetition rates above 100 kHz, where the decreasing expansion time of the annular refractive index structure in the wake of the preceding laser pulse overcomes the µs window for injected light guiding[48,49]. The decrease in the density depletion has been used for triggering electric discharges with lower breakdown potentials for lower gas densities, consistent with Paschen's law[34]. This finding indicates a corresponding lowering of the breakdown threshold required for electric discharge triggering by high-repetition-rate high-average-power laser systems, especially for uncontrollable electric-field conditions as in the laser-lightning rod project[18]. Laser systems with even higher average power at 100 kHz repetition rate are, therefore, promising for achieving long, more depleted channels, which could be of interest for discharge applications and other applications of filamentation.

### Acknowledgments

These results are part of a project that has received funding from the European Research Council (ERC) under the European Union's Horizon 2020 research and innovation programme (grant agreement No. 805202 - Project Teraqua). Funded by the Deutsche Forschungsgemeinschaft (DFG, German Research Foundation) under Germanys Excellence Strategy – EXC-2033 – Projektnummer 390677874 - RESOLV. This project received funding from the European Union's Horizon 2020 research and innovation programme under the Marie Skłodowska-Curie grant agreement No 801459 - FP-RESOMUS. We acknowledge support by the DFG Open Access Publication Funds of the Ruhr-Universität Bochum.

Accepted to APL Photonics 10.1063/5.0175100## Disclosures

DK left the affiliated chair in March 2023 and started to work at a company outside of the field.

## Data Availability Statement

Data underlying the results presented in this paper are available from the authors upon reasonable request.

## References

[1] A. Couairon, and A. Mysyrowicz, "Femtosecond filamentation in transparent media," Phys. Rep. **441**(2–4), 47–189 (2007).

[2] L. Bergé, S. Skupin, R. Nuter, J. Kasparian, and J.-P. Wolf, "Ultrashort filaments of light in weakly ionized, optically transparent media," Rep. Prog. Phys. **70**(10), 1633 (2007).

[3] J.H. Marburger, "Self-focusing: Theory," Prog. Quantum Electron. **4**, 35–110 (1975).

[4] J. Schwarz, P. Rambo, M. Kimmel, and B. Atherton, "Measurement of nonlinear refractive index and ionization rates in air using a wavefront sensor," Opt. Express **20**(8), 8791–8803 (2012).

[5] A. Braun, G. Korn, X. Liu, D. Du, J. Squier, and G. Mourou, "Self-channeling of high-peak-power femtosecond laser pulses in air," Opt. Lett. **20**(1), 73 (1995).

[6] M. Durand, A. Houard, B. Prade, A. Mysyrowicz, A. Durécu, B. Moreau, D. Fleury, O. Vasseur, H. Borchert, K. Diener, R. Schmitt, F. Théberge, M. Chateauneuf, J.-F. Daigle, and J. Dubois, "Kilometer range filamentation," Opt. Express **21**(22), 26836 (2013).

[7] S. Xu, J. Bernhardt, M. Sharifi, W. Liu, and S.L. Chin, "Intensity clamping during laser filamentation by TW level femtosecond laser in air and argon," Laser Phys. **22**(1), 195–202 (2012).

[8] A. Schmitt-Sody, H.G. Kurz, L. Bergé, S. Skupin, and P. Polynkin, "Picosecond laser filamentation in air," New J. Phys. **18**(9), 093005 (2016).

[9] M. Kowalczyk, N. Nagl, P. Steinleitner, N. Karpowicz, V. Pervak, A. Głuszek, A. Hudzikowski, F. Krausz, K.F. Mak, and A. Weigel, "Ultra-CEP-stable single-cycle pulses at 2.2 μm," Optica **10**(6), 801 (2023).

[10] D.S. Steingrube, E. Schulz, T. Binhammer, M.B. Gaarde, A. Couairon, U. Morgner, and M. Kovačev, "High-order harmonic generation directly from a filament," New J. Phys. **13**(4), 043022 (2011).

[11] A. Mysyrowicz, A. Couairon, and U. Keller, "Self-compression of optical laser pulses by filamentation," New J. Phys. **10**(2), 025023 (2008).

[12] L. De La Cruz, E. Schubert, D. Mongin, S. Klingebiel, M. Schultze, T. Metzger, K. Michel, J. Kasparian, and J.-P. Wolf, "High repetition rate ultrashort laser cuts a path through fog," Appl. Phys. Lett. **109**(25), 251105 (2016).

[13] G. Schimmel, T. Produit, D. Mongin, J. Kasparian, and J.-P. Wolf, "Free space laser telecommunication through fog," Optica **5**(10), 1338 (2018).

[14] M.C. Schroeder, I. Larkin, T. Produit, E.W. Rosenthal, H. Milchberg, and J.-P. Wolf, "Molecular quantum wakes for clearing fog," Opt. Express **28**(8), 11463 (2020).

[15] B. Forestier, A. Houard, I. Revel, M. Durand, Y.B. André, B. Prade, A. Jarnac, J. Carbonnel, M. Le Nevé, J.C. De Miscault, B. Esmiller, D. Chapuis, and A. Mysyrowicz, "Triggering, guiding and deviation of long air spark discharges with femtosecond laser filament," AIP Adv. **2**(1), 012151 (2012).

[16] H. Pépin, D. Comtois, F. Vidal, C.Y. Chien, A. Desparois, T.W. Johnston, J.C. Kieffer, B. La Fontaine, F. Martin, F.A.M. Rizk, C. Potvin, P. Couture, H.P. Mercure, A. Bondiou-Clergerie, P. Lalande, and I. Gallimberti, "Triggering and guiding high-voltage large-scale leader discharges with sub-joule ultrashort laser pulses," Phys. Plasmas **8**(5), 2532–2539 (2001).

[17] M. Clerici, Y. Hu, P. Lassonde, C. Milián, A. Couairon, D.N. Christodoulides, Z. Chen, L. Razzari, F. Vidal, F. Légaré, D. Faccio, and R. Morandotti, "Laser-assisted guiding of electric discharges around objects," Sci. Adv. **1**(5), e1400111 (2015).

[18] A. Houard, P. Walch, T. Produit, V. Moreno, B. Mahieu, A. Sunjerga, C. Herkommer, A. Mostajabi, U. Andral, Y.-B. André, M. Lozano, L. Bizet, M.C. Schroeder, G. Schimmel, M. Moret, M. Stanley, W.A. Rison, O. Maurice, B. Esmiller, K. Michel, W. Haas, T. Metzger, M. Rubinstein, F. Rachidi, V. Cooray, A. Mysyrowicz, J. Kasparian, and J.-P. Wolf, "Laser-guided lightning," Nat. Photon., 1–5 (2023).

[19] J. Buldt, H. Stark, M. Müller, C. Grebing, C. Jauregui, and J. Limpert, "Gas-plasma-based generation of broadband terahertz radiation with 640 mW average power," Opt. Lett. **46**(20), 5256 (2021).

[20] T.I. Oh, Y.S. You, N. Jhajj, E.W. Rosenthal, H.M. Milchberg, and K.Y. Kim, "Intense terahertz generation in two-color laser filamentation: energy scaling with terawatt laser systems," New J. Phys. **15**(7), 075002 (2013).

[21] A.D. Koulouklidis, C. Gollner, V. Shumakova, V.Y. Fedorov, A. Pugžlys, A. Baltuška, and S. Tzortzakis, "Observation of extremely efficient terahertz generation from mid-infrared two-color laser filaments," Nat. Commun. **11**(1), 292 (2020).

[22] J.-P. Negel, A. Voss, M.A. Ahmed, D. Bauer, D. Sutter, A. Killi, and T. Graf, "1.1 kW average output power from a thin-disk multipass amplifier for ultrashort laser pulses," Opt. Lett. **38**(24), 5442 (2013).

[23] C. Gaida, M. Gebhardt, T. Heuermann, F. Stutzki, C. Jauregui, and J. Limpert, "Ultrafast thulium fiber laser system emitting more than 1 kW of average power," Opt. Lett. **43**(23), 5853 (2018).

[24] T. Nubbemeyer, M. Kaumanns, M. Ueffing, M. Gorjan, A. Alismail, H. Fattahi, J. Brons, O. Pronin, H.G. Barros, Z. Major, T. Metzger, D. Sutter, and F. Krausz, "1 kW, 200 mJ picosecond thin-disk laser system," Opt. Lett. **42**(7), 1381–1384 (2017).

[25] T. Heuermann, Z. Wang, M. Lenski, M. Gebhardt, C. Gaida, M. Abdelaal, J. Buldt, M. Müller, A. Klenke, and J. Limpert, "Ultrafast Tm-doped fiber laser system delivering

1.65-mJ, sub-100-fs pulses at a 100-kHz repetition rate," Opt. Lett. **47**(12), 3095 (2022).

[26] B.E. Schmidt, A. Hage, T. Mans, F. Légaré, and H.J. Wörner, "Highly stable, 54mJ Yb-InnoSlab laser platform at 0.5kW average power," Opt. Express **25**(15), 17549–17555 (2017).

[27] T. Metzger, A. Schwarz, C.Y. Teisset, D. Sutter, A. Killi, R. Kienberger, and F. Krausz, "High-repetition-rate picosecond pump laser based on a Yb:YAG disk amplifier for optical parametric amplification," Opt. Lett. **34**(14), 2123 (2009).

[28] A.D. Koulouklidis, C. Lanara, C. Daskalaki, V.Y. Fedorov, and S. Tzortzakis, "Impact of gas dynamics on laser filamentation THz sources at high repetition rates," Opt. Lett. **45**(24), 6835–6838 (2020).

[29] A. Higginson, Y. Wang, H. Chi, A. Goffin, I. Larkin, H.M. Milchberg, and J.J. Rocca, "Wake dynamics of air filaments generated by high-energy picosecond laser pulses at 1 kHz repetition rate," Opt. Lett. **46**(21), 5449–5452 (2021).

[30] P. Walch, B. Mahieu, L. Arantchouk, Y.-B. André, A. Mysyrowicz, and A. Houard, "Cumulative air density depletion during high repetition rate filamentation of femtosecond laser pulses: Application to electric discharge triggering," Appl. Phys. Lett. **119**(26), 264101 (2021).

[31] A. Goffin, I. Larkin, A. Tartaro, A. Schweinsberg, A. Valenzuela, E.W. Rosenthal, and H.M. Milchberg, "Optical Guiding in 50-Meter-Scale Air Waveguides," Phys. Rev. X **13**(1), 011006 (2023).

[32] A. Houard, V. Jukna, G. Point, Y.-B. André, S. Klingebiel, M. Schultze, K. Michel, T. Metzger, and A. Mysyrowicz, "Study of filamentation with a high power high repetition rate ps laser at 1.03 μm," Opt. Express **24**(7), 7437–7448 (2016).

[33] E.W. Rosenthal, I. Larkin, A. Goffin, T. Produit, M.C. Schroeder, J.-P. Wolf, and H.M. Milchberg, "Dynamics of the femtosecond laser-triggered spark gap," Opt. Express **28**(17), 24599 (2020).

[34] S. Tzortzakis, B. Prade, M. Franco, A. Mysyrowicz, S. Hüller, and P. Mora, "Femtosecond laser-guided electric discharge in air," Phys. Rev. E **64**(5), 057401 (2001).

[35] T.-J. Wang, M.H. Ebrahim, I. Afxenti, D. Adamou, A.C. Dada, R. Li, Y. Leng, J.-C. Diels, D. Faccio, A. Couairon, C. Milián, and M. Clerici, "Cumulative Effects in 100 kHz Repetition-Rate Laser-Induced Plasma Filaments in Air," Adv. Photonics Res. **4**(3), 2200338 (2023).

[36] A. Talebpour, S. Petit, and S.L. Chin, "Re-focusing during the propagation of a focused femtosecond Ti:Sapphire laser pulse in air," Opt. Commun. **171**(4–6), 285–290 (1999).

[37] D. Abdollahpour, S. Suntsov, D.G. Papazoglou, and S. Tzortzakis, "Measuring easily electron plasma densities in gases produced by ultrashort lasers and filaments," Opt. Express **19**(18), 16866 (2011).

[38] D. Mongin, E. Schubert, L. De La Cruz, N. Berti, J. Kasparian, and J.-P. Wolf, "Linearity of charge measurement in laser filaments," Opt. Express **25**(14), 16517 (2017).

[39] Xin Miao Zhao, J.-C. Diels, Cai Yi Wang, and J.M. Elizondo, "Femtosecond ultraviolet laser pulse induced lightning discharges in gases," IEEE J. Quantum Electron. **31**(3), 599–612 (1995).

[40] P. Polynkin, "Mobilities of O2+ and O2− ions in femtosecond laser filaments in air," Appl. Phys. Lett. **101**(16), 164102 (2012).

[41] S. Henin, Y. Petit, D. Kiselev, J. Kasparian, and J.-P. Wolf, "Contribution of water droplets to charge release by laser filaments in air," Appl. Phys. Lett. **95**(9), 091107 (2009).

[42] Y.-H. Cheng, J.K. Wahlstrand, N. Jhajj, and H.M. Milchberg, "The effect of long timescale gas dynamics on femtosecond filamentation," Opt. Express **21**(4), 4740 (2013).

[43] J.A. Stamper, S.H. Gold, S.P. Obenschain, E.A. McLean, and L. Sica, "Dark-field study of rear-side density structure in laser-accelerated foils," J. Appl. Phys. **52**(11), 6562–6566 (1981).

[44] M. Takeda, H. Ina, and S. Kobayashi, "Fourier-transform method of fringe-pattern analysis for computer-based topography and interferometry," J. Opt. Soc. Am. **72**(1), 156 (1982).

[45] E. Talebian, and M. Talebian, "A general review on the derivation of Clausius–Mossotti relation," Optik **124**(16), 2324–2326 (2013).

[46] J. Chang, D. Li, L. Xu, L. Zhang, T. Xi, and Z. Hao, "Elongation of filamentation and enhancement of supercontinuum generation by a preformed air density hole," Opt. Express **30**(10), 16987 (2022).

[47] F. Paschen, "Ueber die zum Funkenübergang in Luft, Wasserstoff und Kohlensäure bei verschiedenen Drucken erforderliche Potentialdifferenz," Ann. Phys. **273**(5), 69–96 (1889).

[48] O. Lahav, L. Levi, I. Orr, R.A. Nemirovsky, J. Nemirovsky, I. Kaminer, M. Segev, and O. Cohen, "Long-lived waveguides and sound-wave generation by laser filamentation," Phys. Rev. A **90**(2), 021801 (2014).

[49] J.K. Wahlstrand, N. Jhajj, E.W. Rosenthal, S. Zahedpour, and H.M. Milchberg, "Direct imaging of the acoustic waves generated by femtosecond filaments in air," Opt. Lett. **39**(5), 1290 (2014).

List of figures:

Figure 1: Experimental Setup: (HR) high-reflective mirror, (ROC) concave mirror, (BS) beam splitter, (GM) gold mirror, (HV) high voltage. The plate electrodes were moved along the filament by a translation stage. The spherical electrodes had a fixed distance of 30 mm and were placed symmetrically within 1 mm distance to the filament.

Figure 2: Lengths of the plasma channel retrieved by ionic charge measurements (blue) and fluorescence imaging (orange) at 10 kHz laser repetition rate. The saturation of the plasma length at high peak powers has not been observed for 40 kHz and 100 kHz, indicating multi-filamentation at $P \gg P_{cr}$ for the high energy 10 kHz case. A linear correction of the fluorescence-imaging method does not fit the CPP data at all repetition rates.

Figure 3: Comparison of the plasma channel length for constant pulse energies (a) and constant average powers (b).

Figure 4: Ionic charge distribution along the filament exemplary at 10 kHz for different peak powers. The laser is propagating from right to left along the position axis. This data exemplary shows the scaling of the filament length and relative plasma density distribution with peak power from our laser source.

Figure 5: Collected ionic charge per pulse spatially integrated along the filament. The fit shows the linearity of the collected ionic charge.

Figure 6: Modulation of the gas density due to heat deposition, measured by phase-resolved interferometry. The images show the density hole for 5 mJ pulses at a repetition rate of (a) 40 kHz and (b) 100 kHz. The maximum density depletion was measured to be $0.92\rho_0$ and $0.82\rho_0$ for 40 kHz and 100 kHz, respectively. High visibilities of the interference fringes in the raw images of $\geqslant 66\,\%$ provide evidence of the stationarity of the depletion zone.

List of tables:

Table 1: Electric breakdown potentials in kV between identical 30 mm separated spherical electrodes at constant pulse energy and constant average power for laser repetition rates of 10 kHz, 40 kHz, and 100 kHz in air.

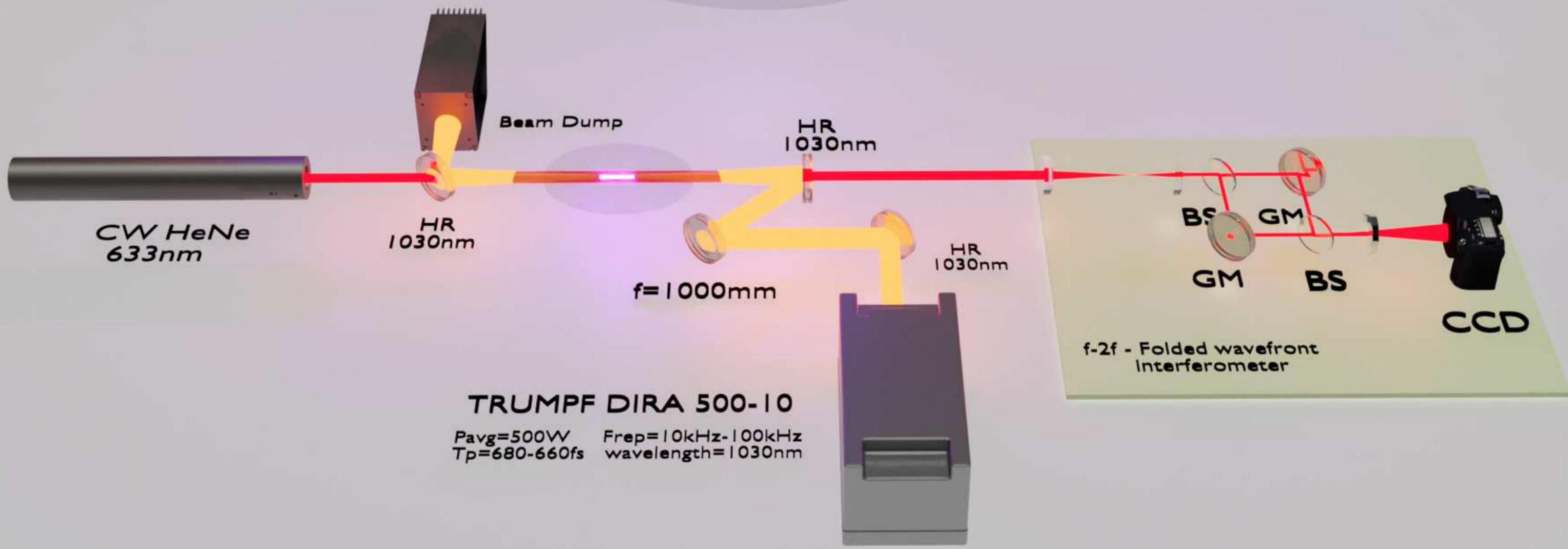

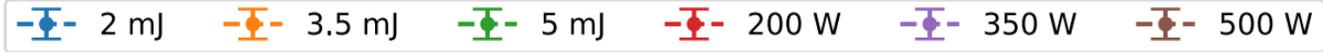
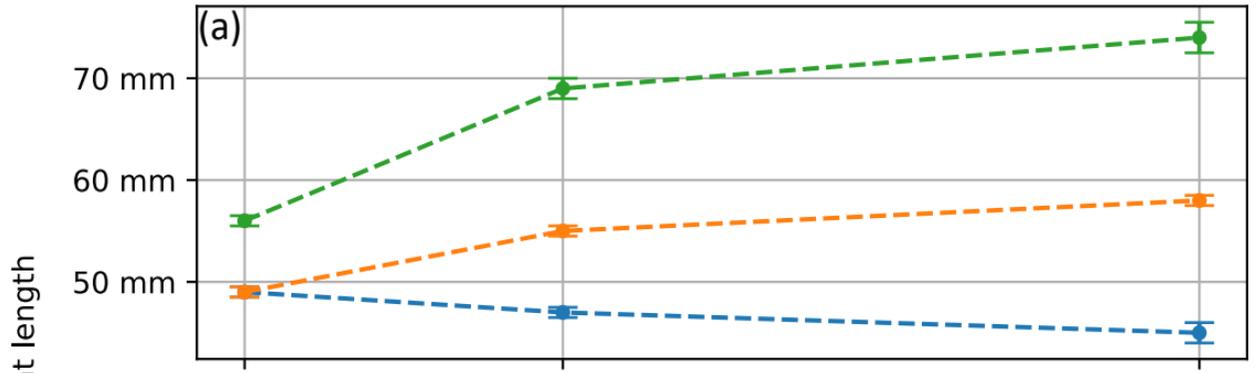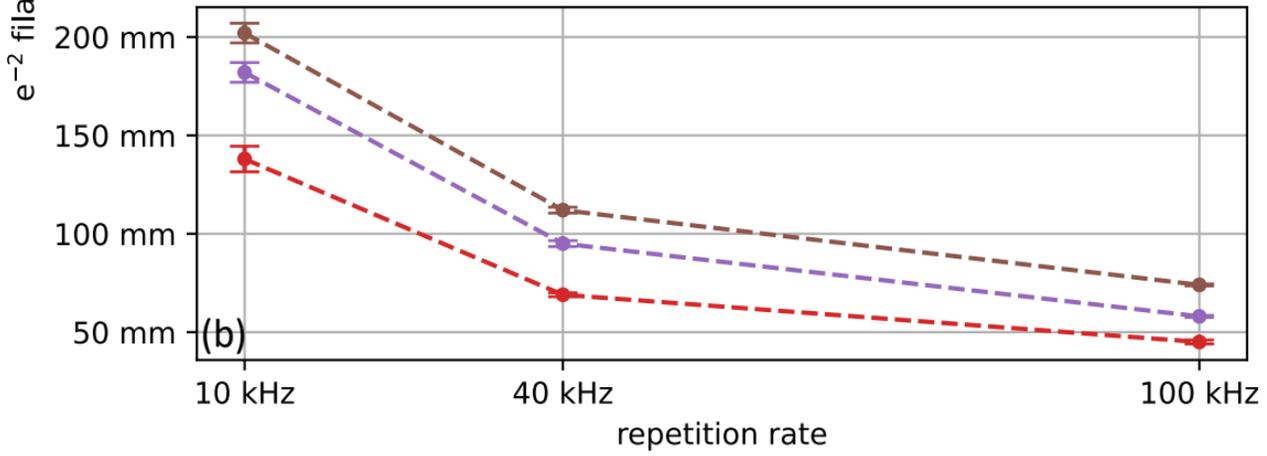

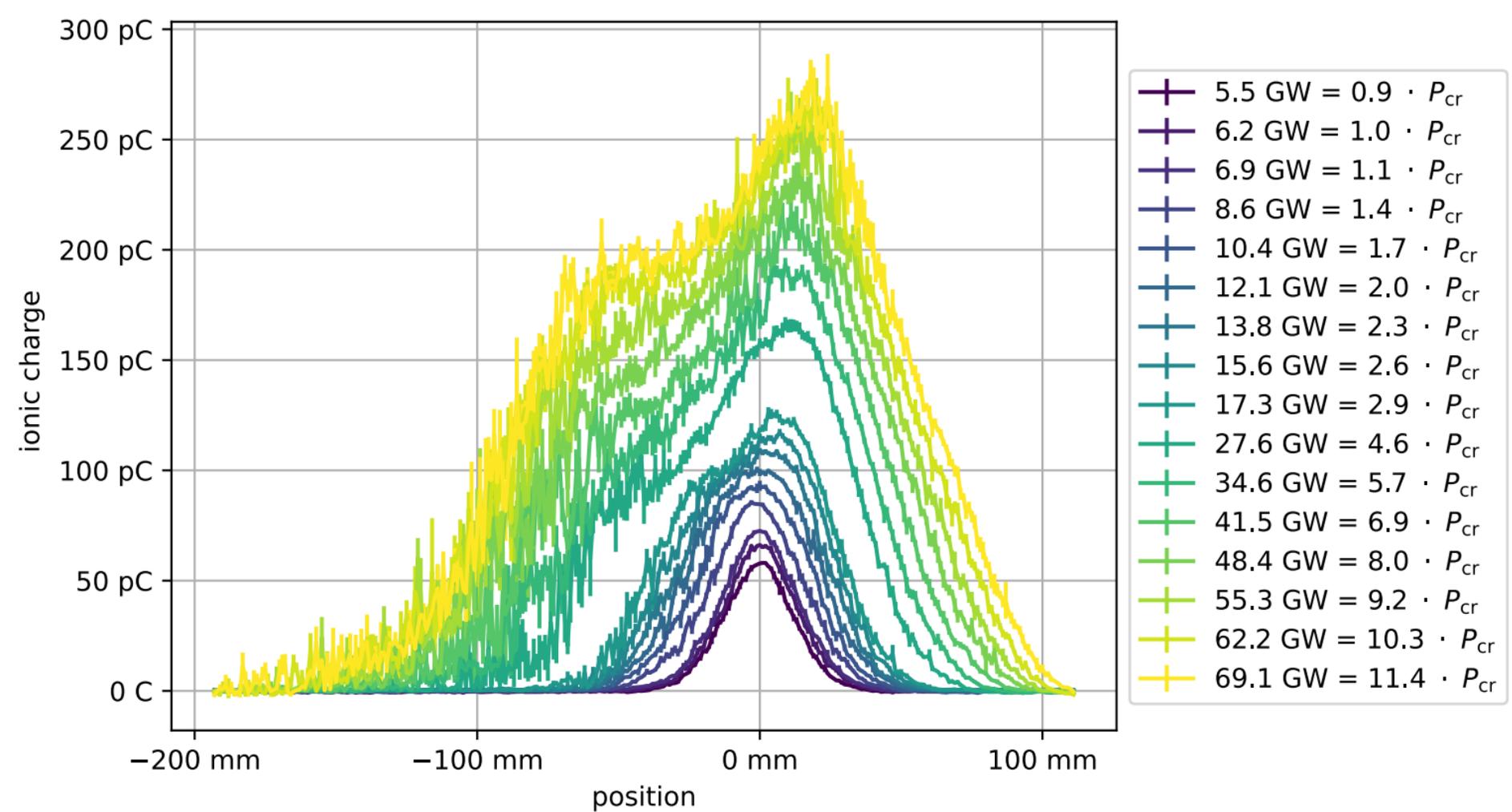

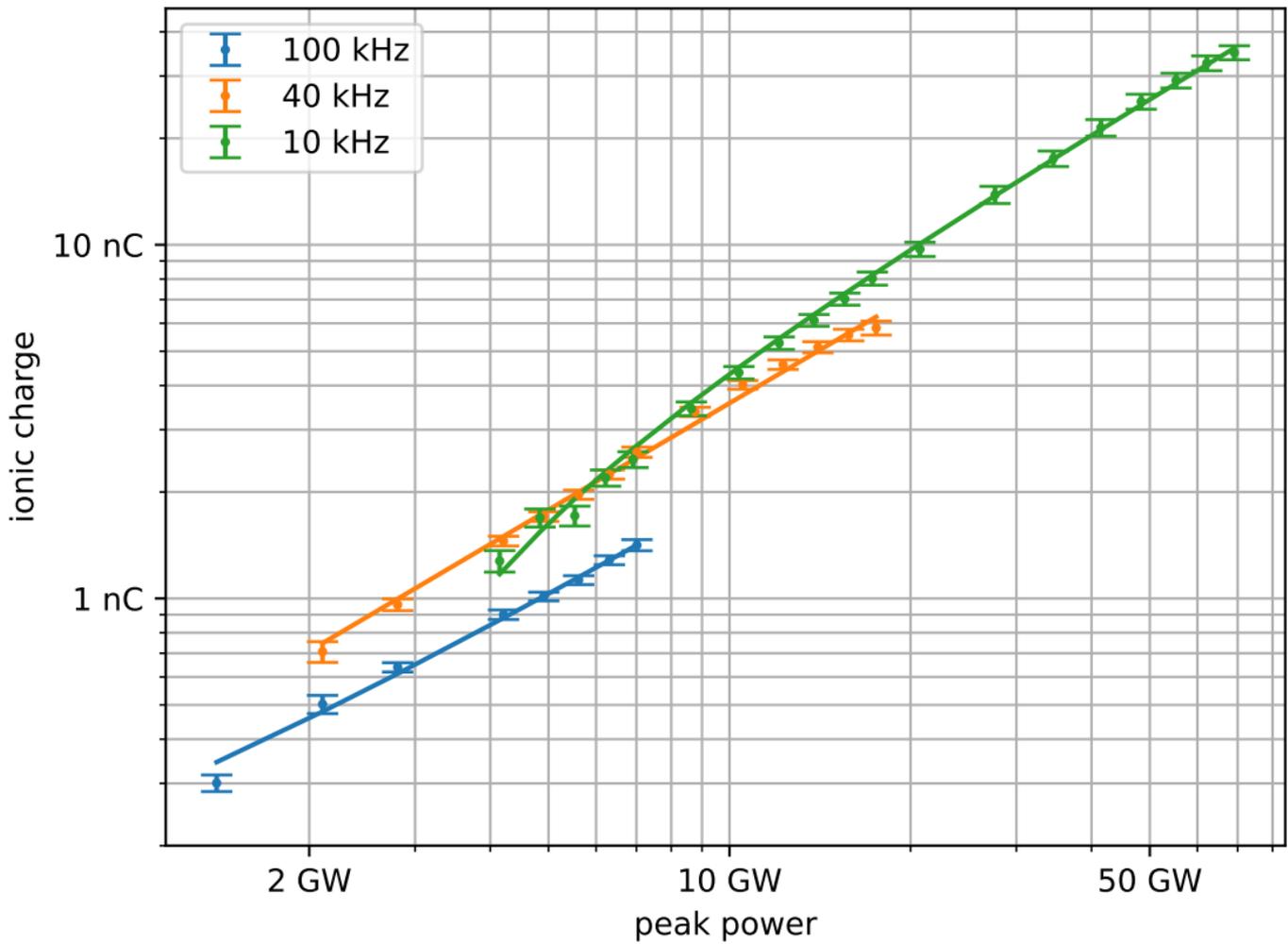

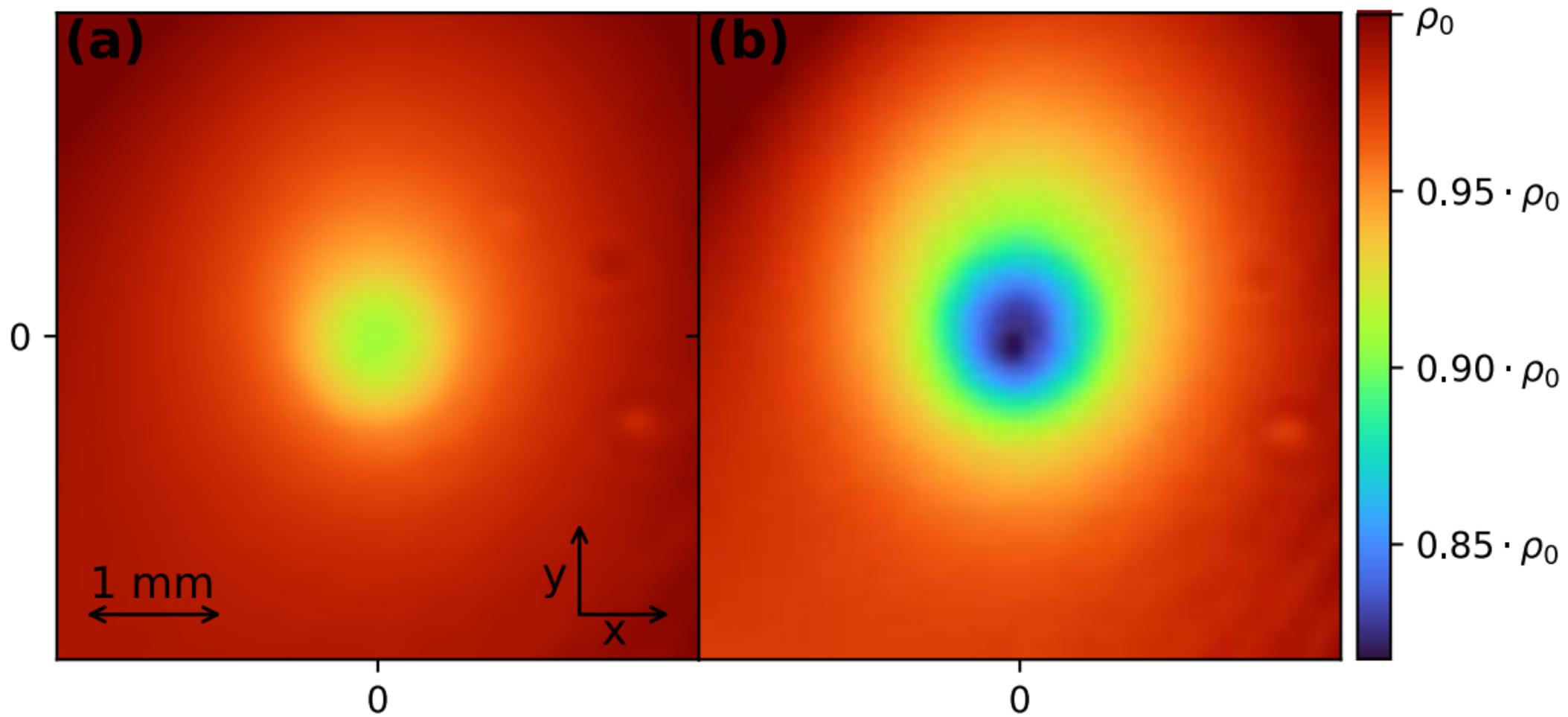